\documentclass[aps,pra,reprint,amsmath,amssymb,showkeys,notitlepage,superscriptaddress]{revtex4-2}
\usepackage{color}
\usepackage{braket}
\usepackage{comment}
\usepackage{graphicx} 
\usepackage{txfonts}  
\usepackage{epstopdf}
\usepackage{appendix}
\usepackage{bigints}
\usepackage{stackengine}
\usepackage{xcolor}
\usepackage{mathtools}
\usepackage{bbold}
\usepackage{braket}
\usepackage{comment}
\newcommand{\md}{\mathrm{d}}
\begin{document}
\title{Boosting energy-time entanglement using coherent time-delayed feedback}

\author{Kisa Barkemeyer}
\thanks{k.barkemeyer@tu-berlin.de}
\affiliation{Institut f\"ur Theoretische Physik, Technische Universit\"at Berlin, D-10623 Berlin, Germany}
\author{Marcel Hohn}
\affiliation{Institut f\"ur Festk\"orperphysik, Technische Universit\"at Berlin, D-10623 Berlin, Germany}
\author{Stephan Reitzenstein}
\affiliation{Institut f\"ur Festk\"orperphysik, Technische Universit\"at Berlin, D-10623 Berlin, Germany}
\author{Alexander Carmele}
\affiliation{Institut f\"ur Theoretische Physik, Technische Universit\"at Berlin, D-10623 Berlin, Germany}

\begin{abstract}
The visibility of the two-photon interference in the Franson interferometer serves as a measure of the energy-time entanglement of the photons. 
We propose to control the visibility of the interference in the second-order coherence function by implementing a coherent time-delayed feedback mechanism. Simulating the non-Markovian dynamics within the matrix product state framework, we find that
the visibility for two photons emitted from a three-level system (3LS) in ladder configuration can be enhanced significantly for a wide range of parameters by slowing down the decay of the upper level of the 3LS. 
\end{abstract}

\maketitle

\section{Introduction}\label{sec:introduction}

Quantum entanglement, according to Schr\"odinger \cite{Schrodinger1935}, is the feature that most clearly distinguishes the quantum from the classical world and has been the subject of ongoing research for decades \cite{Schrodinger1935, Einstein1935, Horodecki2009, Erhard2020}. 
On the one hand, this contributes to our fundamental understanding of quantum mechanics \cite{Einstein1935,Bell1964,Reid2009}. On the other hand, there are various practical fields of application: Entanglement is viewed as a key resource in quantum computation \cite{Bennett1998,Steane1998,OBrien2007}, enables secure quantum communication, for example via quantum key distribution \cite{Gisin2007,Yuan2010}, and facilitates measurements that are more precise than achievable with classical means \cite{Giovannetti2004,Mitchell2004,Giovannetti2011,Slussarenko2017}. 

Photons are well-suited to transmit quantum information as ``flying qubits'' and can be entangled in various degrees of freedom. There has been a lot of research focusing on polarization-entangled photon pairs which, however, suffer from sensitivity to dispersion of the polarization mode when transported in optical fibers \cite{Kwiat1995,Jennewein2000,Muller2014}.
Other types of photon entanglement that have been realized experimentally include position-momentum entanglement \cite{Horne1989} and entanglement in the orbital angular momentum \cite{Mair2001}. 
In 1989, Franson introduced a further type of photon entanglement, namely energy-time entanglement, that is, a high correlation of the photons in their energy and time of emission \cite{Franson1989}. As opposed to polarization entanglement, energy-time entanglement is robust when transported in fibers over long distances \cite{Tittel1998, Inagaki2013}. In connection with the research on energy-time entanglement, also the related time-bin entanglement came into focus \cite{Brendel1999,Marcikic2002,Jayakumar2014,Vedovato2018}.

Franson suggested an ingenious setup to visualize the energy-time entanglement of two photons emitted from a three-level system (3LS) in ladder configuration via an interference in the second-order coherence function \cite{Franson1989}. The Franson interferometer has been realized in a variety of experiments \cite{Ou1990,Brendel1991,Kwiat1993,Aerts1999,Grassani2015,Wakabayashi2015,Agne2017,Peiris2017a,Park2018,Park2019}.
The visibility of the interference fringes depends crucially on the decay rates of the 3LS and a high visibility can be obtained in a parameter regime where the upper state of the 3LS decays slowly in comparison to the middle state.
A typical source of photon pairs are semiconductor quantum dots \cite{Shields2007,Kabuss2011,Su2013,Carmele2019}.
If the photon pairs are generated via a biexciton-exciton cascade, however, the expected visibility is low since, usually, the biexciton state has a shorter lifetime than the exciton state \cite{Moreau2001,Jayakumar2014,Muller2014,Heindel2017,Scholl2020}. 
It is therefore interesting to look for ways to manipulate a 3LS in such a way that the visibility of the interference fringes is increased which indicates an enhanced energy-time entanglement of the emitted photons. While there are resonator-based schemes to deal with this issue, the achievable change of the emission dynamics is essentially given and limited by the device geometry \cite{Bayer2001,Gevaux2006,Jakubczyk2014,Scholl2020}.
Complementarily, time-delayed feedback has proven to be a versatile tool for the control of classical as well as quantum systems \cite{Pyragas1992,Bechhoefer2005,Scholl2016}. In quantum systems, measurement-based feedback control schemes are detrimental to entanglement since during the measurement process quantum coherence is destroyed \cite{Wiseman1994, Sayrin2011,Rossi2018}. 
In contrast, coherent time-delayed feedback preserves quantum coherence and, thus, is well suited for the control of entanglement \cite{Lloyd2000,Carmele2013,Tufarelli2013,Hein2014,Tufarelli2014,Fang2015,Grimsmo2015,Guimond2017,Barkemeyer2019,Calajo2019a,Sinha2020,Katsch2020,Barkemeyer2020}. 

Here, we propose a scheme to control the visibility of the two-photon interference in the Franson interferometer via the manipulation of the decay rates in the 3LS using coherent time-delayed feedback. We treat the problem within the matrix product state (MPS) framework which allows the numerically exact simulation of the non-Markovian system dynamics \cite{Pichler2016,Guimond2017}. 

The paper is structured as follows: After this introduction in Sec.~\ref{sec:introduction}, in Sec.~\ref{sec:Model} we present our model and the MPS method we use to calculate the time evolution and to evaluate the second-order coherence function numerically. Furthermore, we benchmark the results we obtain in the case without feedback where the problem can be solved analytically. Subsequently, in Sec.~\ref{sec:results} we present the results we obtain when a feedback mechanism is implemented and finally, in Sec.~\ref{sec:Conclusion}, we summarize our findings.

\section{Model}
\label{sec:Model}

\begin{figure}
    \centering
    \includegraphics[]{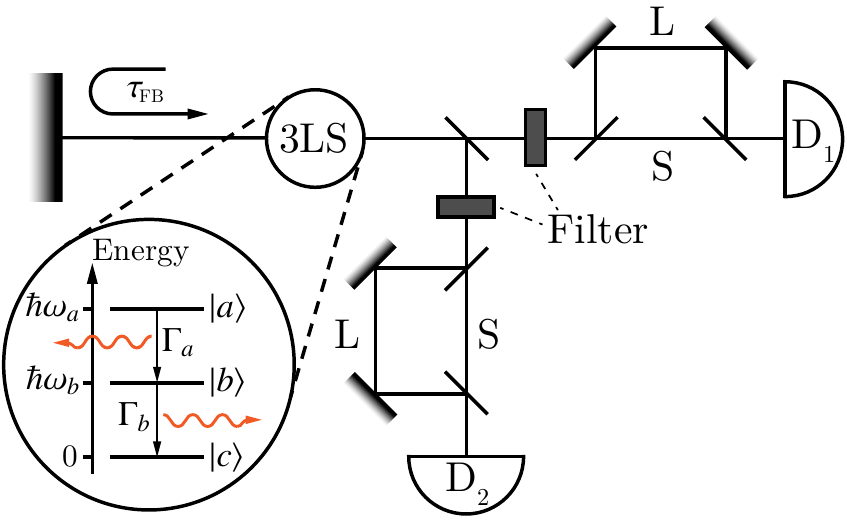}
    \caption{(Color online) Franson interferometer with feedback channel. The setup allows to measure the second-order coherence function of two photons emitted from a three-level system (3LS) in ladder configuration. For more pronounced effects, the emitted photons are frequency filtered which ensures that the first photon is detected by the first detector, $D_1$, and the second photon is detected by the second detector, $D_2$, if a signal is registered. 
    The photons can reach the detectors either via short ($S$) or long paths ($L$) defined by beamsplitters and mirrors. For the $\ket{a} \leftrightarrow \ket{b}$ transition, a feedback mechanism at delay time $\tau_\text{FB}$ is implemented.}
    \label{fig:FransonSetup}
\end{figure}

The setup we consider is illustrated in Fig.~\ref{fig:FransonSetup}. The Franson interferometer allows the analysis of two-photon correlations. We assume the source of these photons to be a ladder-type 3LS initially in the upper state $\ket{a}$ which decays with rate $\Gamma_a$ to the middle state $\ket{b}$ emitting a photon. The middle state, in turn, decays with rate $\Gamma_b$ to the ground state $\ket{c}$ under the emission of a second photon.
The photons emitted from the 3LS impinge on a 50:50 beamsplitter and enter into one of the two unbalanced Mach-Zehnder interferometer arms constituting the Franson interferometer. In each of these arms, the photons can take either a short path of length $S$ or a long path of length $L$ defined by beamsplitters and mirrors to the respective detector. Note that only those events are registered in which the two photons enter into different arms of the interferometer which corresponds to postselection. Additionally, by placing frequency filters in the optical paths we ensure that we know which photon has entered into which channel of the interferometer if a signal is registered. 
In principle, it is also possible to not differentiate between the photons in which case they are both affected by the feedback mechanism we implement. 
For the sake of simplicity and numerical convenience, however, we focus on the given setup where we are able to control the photons individually. Here, the effects that occur are more pronounced and serve as a proof of principle.

When measuring the second-order coherence function, the interference of the two-photon probability amplitudes of the photons taking either the short or the long paths results in an oscillation of the height of the central peak with the relative phase between the paths.  This interference can be observed in the parameter regime where the delay time between the long and the short path to the detectors, $T = (L-S)/c$ with $c$ being the speed of light in the waveguide, exceeds the first-order coherence time so that single-photon interference effects can be ruled out.
The visibility of this interference points to the energy-time entanglement of the photons.

In the setup we consider here, we subject the 3LS to coherent time-delayed feedback which allows us to control the visibility of the interference of the second-order coherence function. Concretely, we implement a feedback mechanism at delay time $\tau_\text{FB}$ for the first decay channel, that is, for the photon that is emitted first from the 3LS.

The Hamiltonian governing the dynamics in the rotating wave and dipole approximation reads
\begin{multline}
\mathcal{H} = \hbar \omega_a \sigma_{aa} + \hbar \omega_b \sigma_{bb} + \hbar \int \mathrm{d}\omega \omega \left(r^{\dagger(1)}_\omega r^{(1)}_\omega + r^{\dagger(2)}_\omega r^{(2)}_\omega\right) \\
+ \hbar \int \mathrm{d}\omega g_a(\omega) \left(\sigma_+^{(1)}r_\omega^{(1)} +\text{H.c.} \right) + \hbar \int \mathrm{d}\omega g_b(\omega) \left(\sigma_+^{(2)}r_\omega^{(2)} +\text{H.c.} \right).
\label{eq:Ham}
\end{multline}
The occupation of the upper (middle) level of the 3LS with energy $\hbar \omega_a$ ($\hbar \omega_b$) is described by the operator $\sigma_{aa} = \ket{a}\bra{a}$ ($\sigma_{bb} = \ket{b}\bra{b}$) while the energy of the ground state of the 3LS is set to zero. The annihilation (creation) of a photon with energy $\hbar \omega$ can be described by the bosonic operator $r_\omega^{(\dagger)(i)}$, $i \in \{1,2\}$, where we consider two separate reservoirs, namely the two arms of the interferometer. The second line of Eq.~\eqref{eq:Ham} arises due to the coupling between the 3LS and these reservoirs. The $\ket{a} \leftrightarrow \ket{b}$ ($\ket{b} \leftrightarrow \ket{c}$) transition, encoded in the transition operator $\sigma^{(1)}_+ = \ket{a}\bra{b}$ ($\sigma^{(2)}_+ = \ket{b}\bra{c}$), couples to the first (second) reservoir with coupling strength $g_a(\omega)$ ($g_b(\omega)$) which is, in general, frequency-dependent. If we implement a feedback mechanism with delay time $\tau_\text{FB}$ for the first channel only, the coupling strength for this channel takes on a sinusoidal form, $g_a(\omega) = g_a \sin\left(\omega \tau_\text{FB}/2\right)$, while the coupling strength for the second channel can be assumed to be constant, $g_b(\omega) = g_b$.

In the rotating frame defined by its freely-evolving part, that is, the first line of Eq.~\eqref{eq:Ham}, the Hamiltonian takes on the form 
\begin{multline}
    \mathcal{H}'(t) = \hbar \int \mathrm{d}\omega g_a(\omega) \left(\sigma_+^{(1)}r_\omega^{(1)}e^{i(\omega_{ab}-\omega)t} +\text{H.c.} \right) \\
    + \hbar \int \mathrm{d}\omega g_b(\omega) \left(\sigma_+^{(2)}r_\omega^{(2)}e^{i(\omega_b-\omega)t} +\text{H.c.} \right)
    \label{eq:Ham_rot}
\end{multline}
with $\omega_{ab} \equiv \omega_a -\omega_b$ being the transition frequency of the $\ket{a} \leftrightarrow \ket{b}$ transition.

We are interested in the second-order coherence function of the photons which is defined as
\begin{align}
    G^{(2)}(\tau) &= \int_0^\infty \mathrm{d}t_1 \left.G^{(2)}(t_1,t_2)  \right|_{t_2 \rightarrow t_1 + \tau}, \label{eq:G2tau}\\
G^{(2)}(t_1,t_2) &= \bra{\Psi} E^{(-)}_1(t_1) E^{(-)}_2(t_2) E^{(+)}_2(t_2) E^{(+)}_1(t_1) \ket{\Psi}.
\label{eq:G2t1t2}
\end{align}
Here, $\ket{\Psi}$ represents the two-photon state emitted from the 3LS. The operator $E_i^{(+)}$ is the positive frequency part of the electric field operator at detector $i$, $E_i$, and comprises the photon annihilation operators. As its hermitian conjugate, $E_i^{(-)}$ is the negative frequency part so that
\begin{equation}
    E_i(t) = E_i^{(+)}(t) + E_i^{(-)}(t).
\end{equation}
To account for the two equiprobable paths the photons can take to the detectors, we split up the operators according to
\begin{equation}
    E^{(+)}_i(t) = \frac12 \left[ E^{(+)}_{i,S}(t) + E^{(+)}_{i,L}(t) \right]
    \label{eq:splitSL}
\end{equation}
where the first term on the right-hand side refers to photon $i$ taking the short, the second term to the photon taking the long path to detector $i$.

\subsection{Time evolution in the MPS framework}

\begin{figure*}
    \centering
    \includegraphics[]{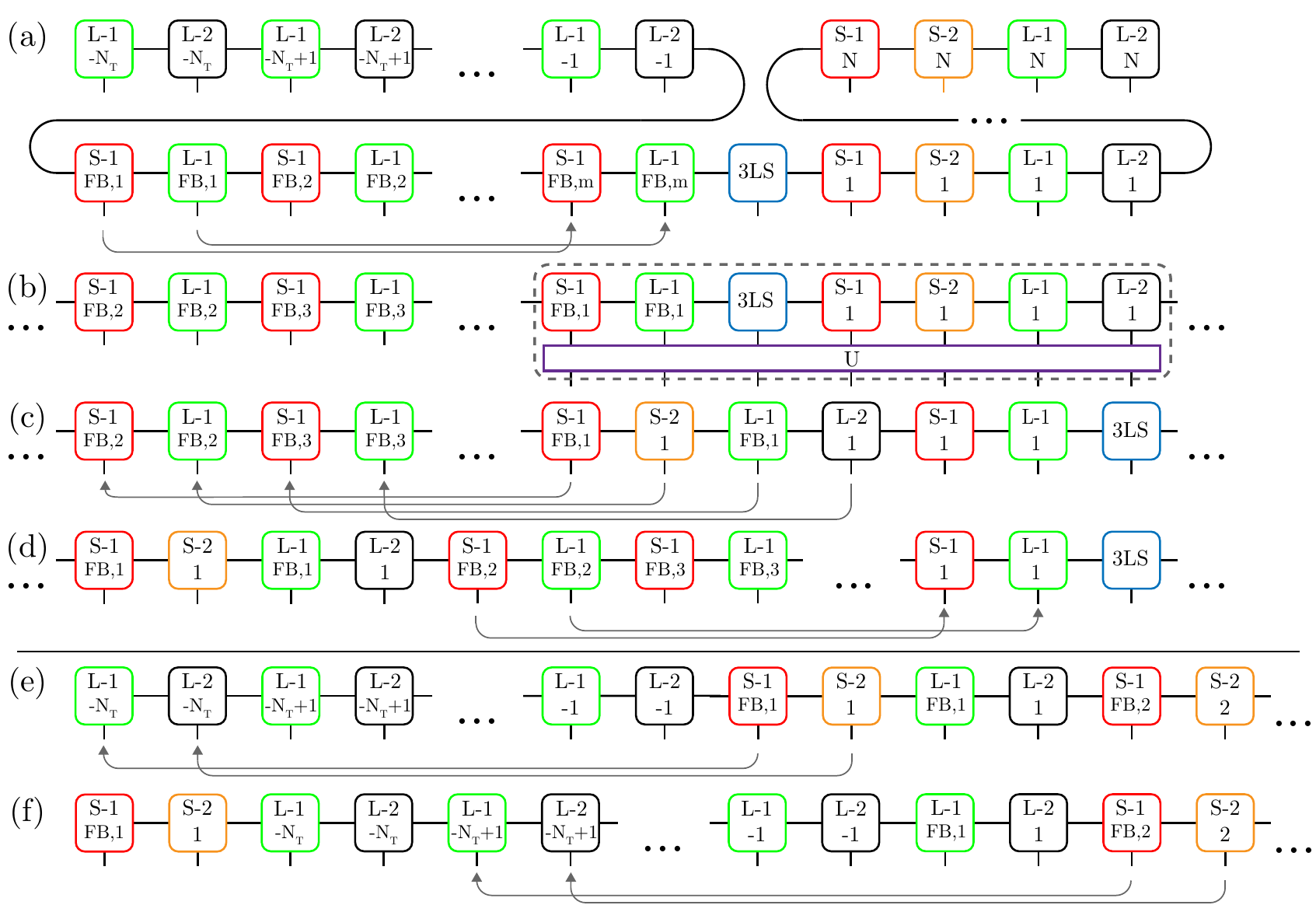}
    \caption{(Color online) Tensor diagram illustrating the time evolution algorithm. There is a bin describing the three-level system (3LS, blue box) and four types of time bins describing the short path to the first detector (S-1, red box), the short path to the second detector (S-2, orange box), the long path to the first detector (L-1), and the long path to the second detector (L-2, black box). (a) The time bins containing the feedback signal are swapped next to the 3LS bin. (b) The feedback bins, the 3LS bin and the current time bins are contracted with the stroboscopic time evolution operator (U, purple box). (c) The output bins are placed to the left of the bins in the feedback channel. (d) The procedure is repeated $N$ times. (e) After the time evolution is completed, the bins referring to the short paths are swapped $2N_T$ steps to the left and subjected to a phase shift to account for the delay time between short and long paths. (f) The procedure is repeated $N$ times.}
    \label{fig:Tensordiagram_timeevo}
\end{figure*}

Here, we introduce the time evolution algorithm based on MPS in a condensed form and refer the interested reader to Refs.~\cite{Pichler2016} for further details. The numerical calculations in this paper were performed using the ITensor Library \cite{itensor}.
We model the system within the MPS framework where the continuous dynamics is discretized into time steps $\Delta t$ which are small compared to the time scales of the system evolution. 
In the case of feedback with delay time $\tau_\text{FB}=m\Delta t$, $m \in \mathbb{N}$, on the first channel, the evolution from time step $k$ to $k+1$, $k \in \mathbb{N}$, $t_{k+1}-t_k = \Delta t$, can be realized on the basis of the Schr\"odinger equation via the stroboscopic time evolution operator $U_k$ as the dynamical map
\begin{equation}
    \ket{\psi(t_{k+1})} = U_k \ket{\psi(t_k)}.
\end{equation}
As derived in detail in the Appendix~\ref{Ap:0}, $U_k$ takes the form
\begin{multline}
    U_k = \exp\left\{ \sqrt{\Gamma_a} \left[\sigma_+^{(1)} \left(\Delta R^{(1)}_k - \Delta R^{(1)}_{k-m}e^{i \omega_{ab} \tau_\text{FB}} \right) -\text{H.c.}\right] \right.  \\
    \left.-i \sqrt{\Gamma_b} \left[\sigma_+^{(2)} \Delta R^{(2)}_k + \text{H.c.}\right] \right\}
\end{multline}
with the decay rates $\Gamma_a =  g_a^2 \pi/2$ and $\Gamma_b = 2 \pi g_b^2 $. Note that the definitions of the decay rates differ by a factor so that the Markovian channel without feedback and the non-Markovian feedback channel can be treated on equal footing without having to take additional time bins for the Markovian channel into consideration.
The creation of a photon in time bin $k$ of reservoir $i$, $i \in \{ 1,2\}$, is described by the noise increment
\begin{equation}
    \Delta R_k^{\dagger (i)} = \int_{t_k}^{t_{k+1}}\md t r^{\dagger (i)}_t
\end{equation}
where the time-dependent noise operator $r_t^{\dagger(i)}$ models the creation of a photon at time $t$ as the fourier transform of the operator $r_\omega^{\dagger (i)}$ referring to the creation of a photon with frequency $\omega$,
\begin{equation}
    r^{\dagger (i)}_t = \frac{1}{\sqrt{2\pi}}\int \md \omega r_\omega^{\dagger (i)} e^{i\left(\omega-\omega_i\right)t},
    \label{eq:noise_op}
\end{equation}
with $\omega_1 = \omega_{ab}$, $\omega_2 = \omega_b$.

The noise increments obey $\left[\Delta R_k^{(i)}, \Delta R_{k'}^{\dagger (i')}\right] = \Delta t \delta_{kk'} \delta_{ii'}$ and define a discrete, othonormal time-bin basis of the Hilbert space. With the multi-index $l \equiv (i,k)$ denoting time step $k$ of channel $i$, the Fock state of time bin $l$ containing $j_l$ photons, $j_l \in \mathbb{N}$, can be created via
\begin{equation}
    \ket{j_l}_l = \frac{\left[ \Delta R_k^{\dagger (i)}\right]^{j_l}}{\sqrt{j_l!}\left(\Delta t\right)^{j_l}}\ket{\text{vac}}_l.
\end{equation}
In this basis, the combined state of the 3LS and the photonic reservoirs takes the form
\begin{equation}
    \ket{\psi(t_k)} = \sum_{j_S,j_1,\dots,j_N} \Psi_{j_S,j_1,\dots,j_N}  \ket{j_S,j_1,\dots,j_M}
\end{equation}
where the index $j_S \in \{a,b,c\}$ describes the state of the 3LS while the index $j_l$, $l \in \{1,\dots,M\}$ denotes the occupation of the $l$-th of $M$ reservoir time bins.
In general, the dimension of the Hilbert space grows exponentially with the number of time bins since the complex coefficient tensor $\Psi$ is $3p^M$ dimensional where $(p-1)$ is the maximum number of photons per time bin considered. The key idea of the time evolution algorithm based on MPS is to decompose the coefficient tensor into a product of matrices via a series of singular value decompositions,
\begin{equation}
    \Psi_{j_S,j_1,\dots,j_M} = A^{j_S}A^{j_1}\cdots A^{j_M}.
\end{equation}
This decomposition allows an efficient treatment of the entangled state of the system and the reservoir 
by neglecting the smallest singular values which corresponds to a justified truncation of the least entangled and, thus, least important parts of the Hilbert space.
Since each matrix $A$ contains a single physical index $j_{S/l}$, furthermore, a time-local description is obtained so that for each time step, the time evolution operator only has to be applied to the 3LS bin and the involved time bins. To efficiently include the feedback contributions which correspond to long-range interactions in time, a swapping procedure is performed shifting the feedback bins next to the 3LS bin.

In our system, analogously to Eq.~\eqref{eq:splitSL}, we split up the noise increments to account for the two possible paths to the detectors according to
\begin{equation}
    \Delta R^{\dagger (i)}_k = \frac12 \left(\Delta R^{\dagger (i)}_{k,S} + \Delta R^{\dagger (i)}_{k,L} \right).
\end{equation}
The delay time between the long and the short path to the detectors, $T$, corresponds to $N_T = T/\Delta t$ time bins.

The time evolution algorithm is illustrated in Fig.~\ref{fig:Tensordiagram_timeevo}. 
For each of the reservoirs we use two time bins per time step to account for the two possible paths to the detectors so that, in total, there are four time bins per time step: S-1 (short path to first detector, red box), S-2 (short path to second detector, orange box), L-1 (long path to first detector, green box), and L-2 (long path to second detector, black box). 
At the beginning of each time step, the bins containing the feedback signal (S-1 and L-1) are swapped $2(m-1)$ steps to the right and placed next to the 3LS bin (blue box) (a) so that the time evolution can be performed efficiently by contracting the 3LS bin, the feedback bins, the current time bins and the time evolution operator (U, purple box) (b). Subsequently, the contracted tensor is decomposed and the four output bins are swapped to the left of the bins in the feedback channel while the two time bins that are kept next to the 3LS bin account for the emission into this feedback channel (c). Afterwards, the next time evolution step can be performed (d). After $N$ time steps, the time evolution is completed and we rearrange the bins to account for the different path lengths. To that end, all bins describing the photons taking the short paths to the detectors are swapped $2 N_T$ steps to the left and subjected to a phase shift to account for the relative phase of the photons taking the short and the long paths (e). At the beginning of the MPS, we place vacuum bins which describe the initial vacuum signal reaching the detectors via the long paths. The step is repeated (f) until all bins referring to the short paths have been moved. 
For simplicity, we assume that the relative phase between the short and the long paths has the same effect in both arms of the interferometer, that is,
\begin{equation}
\phi_T \equiv \omega_{ab} T = \omega_b T + 2\pi z
\label{eq:assume_phaseT}
\end{equation}
where $z \in \mathbb{Z}$. If this was not the case, we could either consider divergent phases or adjust the phases via an asymmetric interferometer with different delay times $T_a$ and $T_b$ in the two arms.

\subsection{Evaluation of the second-order coherence function}

\begin{figure}
    \centering
    \includegraphics[]{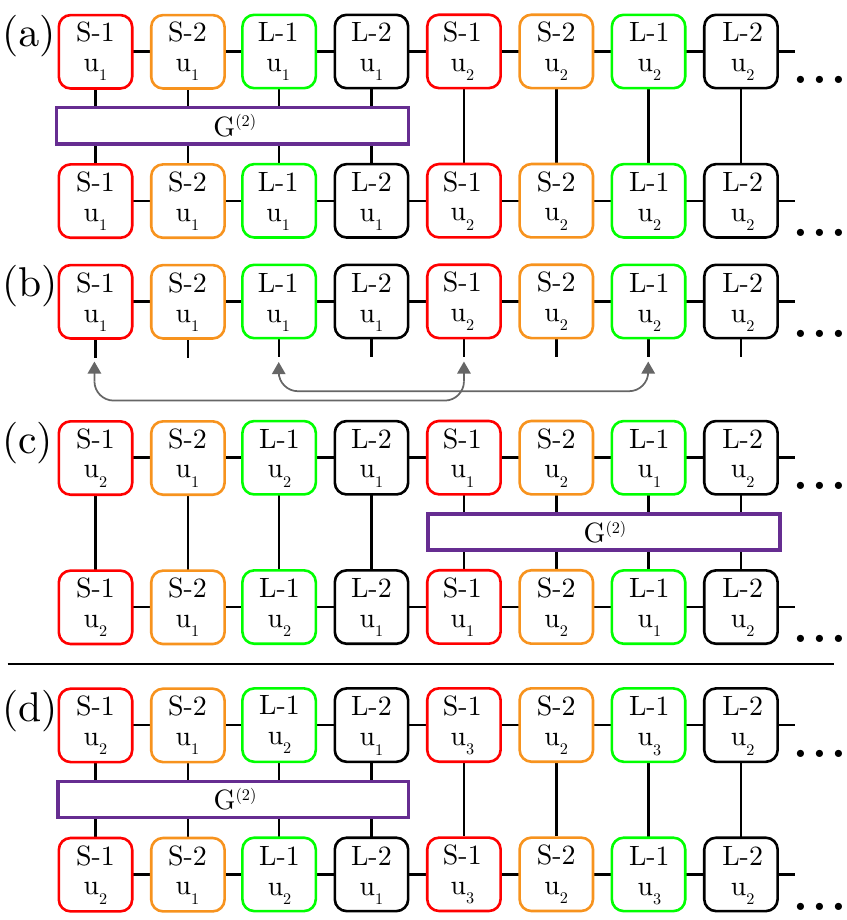}
    \caption{(Color online) Tensor diagram illustrating the evaluation of the two-time second-order coherence function $G^{(2)}(t_1,t_2)$, $t_1, t_2 \in \{u_1, \dots, u_N\}$ where two nested loops are necessary to account for all possible values of $t_1$ and $t_2$. (a) Evaluation for $t_1 = t_2 =u_1$. (b) The time bins containing the signal registered at detector $D_1$, S-1 and L-1, for time step $u_1$ are exchanged with the corresponding time bins for the next time step $u_2$. (c) Evaluation of $G^{(2)}(u_1,u_2)$. The procedure is repeated until $G^{(2)}(u_1,u_k)$ has been evaluated for all $k \in \{1,\dots,N\}$. (d) Afterwards, we start the second loop through the MPS by calculating $G^{(2)}(u_2,u_1)$ and so on until we have completely determined $G^{(2)}(t_1,t_2)$.}
    \label{fig:Tensordiagram_g2}
\end{figure}

After the time evolution and the time and phase shift accounting for the different path lengths to the detectors have been performed, we can evaluate the two-time second-order coherence function $G^{(2)}(t_1,t_2) = \bra{\Psi}E_1^{(-)}(t_1)E_2^{(-)}(t_2)E_2^{(+)}(t_2)E_1^{(+)}(t_1)\ket{\Psi}$ introduced in Eq.~\eqref{eq:G2t1t2} via the algorithm illustrated in Fig.~\ref{fig:Tensordiagram_g2}. The output bins describe the signal that reaches the detectors at a certain time step $u_i$, $i \in \{1,\dots,N\}$. 
We start at the beginning of the MPS and calculate $G^{(2)}(u_1,u_1)$ (a). Afterwards, we exchange the time bins describing the signal at the first detector, S-1 and L-1, at this time step, $u_1$, with the S-1 and L-1 bins of the next time step, $u_2$, (b) so that we can efficiently evaluate $G^{(2)}(u_1,u_2)$ (c). This step is repeated until $G^{(2)}(u_1,u_k)$ has been evaluated for all $k \in \{1,\dots,N\}$. Since we need the two-time second-order coherence function $G^{(2)}(t_1,t_2)$ for all possible combinations of $t_1$ and $t_2$, we then start again at the beginning evaluating $G^{(2)}(u_2,u_1)$ and so on until $G^{(2)}(t_1,t_2)$ has been completely determined. By performing an integration, see Eq.~\eqref{eq:G2tau}, we subsequently obtain the second-order coherence as a function of the delay between the detection of the first and the second photon, $G^{(2)}(\tau)$. 

\subsection{Benchmark without feedback}

To ensure the validity of our approach we benchmark the MPS results in the special case without feedback where an analytical solution is possible. 
Numerically, we realize this scenario by omitting the feedback channel and performing the time evolution using the time evolution operator we obtain in the case $g_a(\omega)=g_a$ and $g_b(\omega)=g_b$,
\begin{multline}
U_k^{\text{no FB}} =  \exp\left\{ -i\sqrt{\Gamma_a} \left[\sigma_+^{(1)} \Delta R^{(1)}_k  +\text{H.c.}\right]\right. \\
\left. -i \sqrt{\Gamma_b} \left[\sigma_+^{(2)} \Delta R^{(2)}_k + \text{H.c.}\right] \right\}
\end{multline}
with $\Gamma_a = 2\pi g_a^2$, $\Gamma_b = 2\pi g_b^2$.
For the analytical solution which is based on Refs.~\cite{Meyer1997,Scully1997} we need an expression for the two-photon state that is emitted from the 3LS which, as derived in the Appendix~\ref{Ap:1}, takes the form 
\begin{multline}
 \ket{\Psi} = \hspace{-0.5em} \bigintsss\hspace{-1em}\bigintsss \hspace{-0.5em} \mathrm{d}\omega \mathrm{d}\omega' \\ \times \frac{ - g_a g_b}{\left[i \left(\omega +\omega'- \omega_{a} \right) - \frac{\Gamma_a}{2} \right]\left[i \left( \omega'-\omega_{b}\right) -\frac{\Gamma_b}{2} \right]} \ket{1_{\omega}, 1_{\omega'}}. \label{eq:two-photState}
\end{multline}

\begin{figure}
\centering
\includegraphics[]{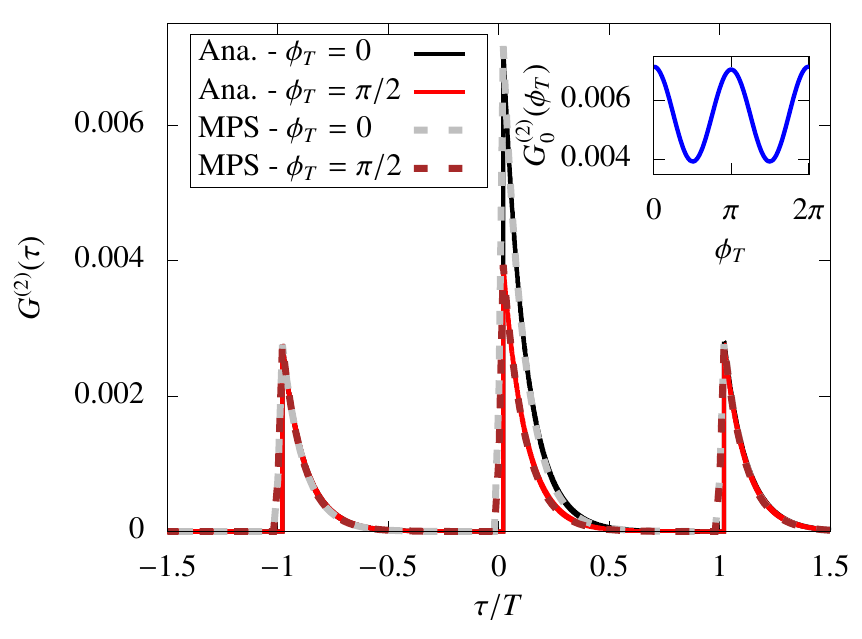}
\caption{(Color online)  Benchmark of the MPS solution for the second-order coherence function $G^{(2)}$ without feedback. The MPS solution is compared to the analytical solution (Ana.) for different values of the relative phase $\phi_T$ between the short and the long paths to the detectors. Here, $\Gamma_a T = 2.5$, $\Gamma_b T = 10$, $\Delta t = 0.05/\Gamma_a$, $\Gamma_a t_\text{max}= 10$. (Inset) Oscillation of the height of the central peak with the delay phase $\phi_T$.}
\label{fig:Benchmark}
\end{figure}

\begin{figure}
\centering
    \includegraphics[]{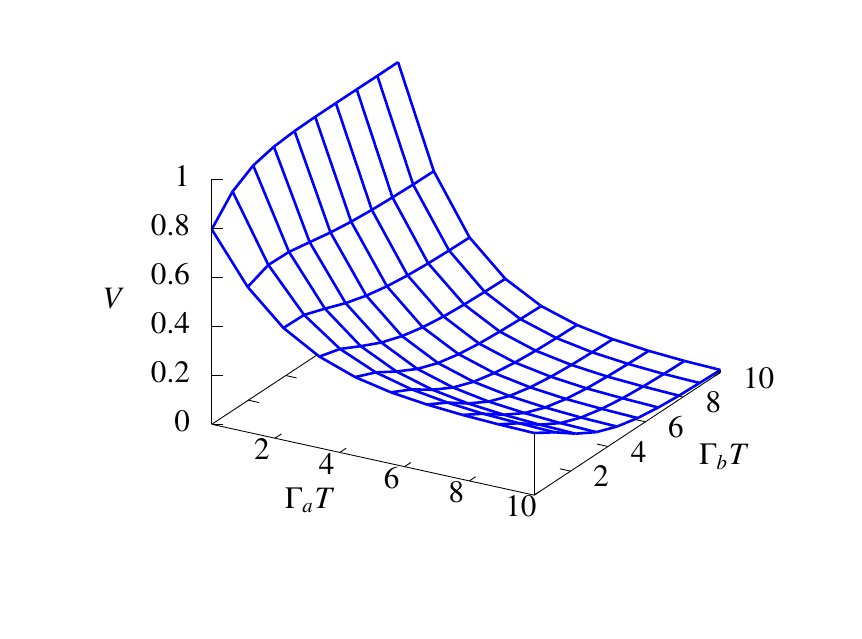}
\caption{(Color online) Visibility of the $G^{(2)}$ interference for the Franson interferometer without feedback as a function of $\Gamma_a T$ and $\Gamma_b T$ calculated analytically.}
\label{fig:vis_noFB}
\end{figure}

We rewrite the two-time second-order coherence function $G^{(2)}(t_1,t_2)$ from Eq.~\eqref{eq:G2t1t2} introducing the two-photon probability amplitude $\Psi(t_1,t_2)$ as
\begin{equation}
    G^{(2)}(t_1,t_2) = \left|\Psi(t_1,t_2)\right|^2,
\end{equation}
\begin{align}
    &\Psi(t_1,t_2)\notag\\
    &= \frac{1}{4}\bra{0}\left[E^{(+)}_{2,S}(t_2) + E^{(+)}_{2,L}(t_2) \right]\left[E^{(+)}_{1,S}(t_1) + E^{(+)}_{1,L}(t_1)\right]\ket{\Psi} \notag \\
    &= \Psi_{S,S}(t_1,t_2) + \Psi_{S,L}(t_1,t_2) + \Psi_{L,S}(t_1,t_2) + \Psi_{L,L}(t_1,t_2)
\end{align}
where
\begin{equation}
    \Psi_{r_1,r_2}(t_1,t_2) = \bra{0}E^{(+)}_{2,r_2}(t_2)E^{(+)}_{1,r_1}(t_1)\ket{\Psi}
    \label{eq:two-photAmpl}
\end{equation}
is the two-photon probability amplitude describing the first photon reaching the first detector via the path $r_1$ and the second photon reaching the second detector via the path $r_2$ with $r_1,r_2 \in \{S,L\}$.
As shown in the Appendix~\ref{Ap:2}, the evaluation of Eq.~\eqref{eq:two-photAmpl} using the two-photon state $\ket{\Psi}$ from Eq.~\eqref{eq:two-photState} results in
\begin{multline}
\Psi_{r_1,r_2}(t_1,t_2) =  \eta e^{-\left( i\omega_{a}+\frac{\Gamma_a}{2}\right)\left(t_1 - \frac{r_1}{c} \right)} \Theta\left(t_1 - \frac{r_1}{c} \right) \\
\times e^{-\left(i\omega_{b} + \frac{\Gamma_b}{2}\right)\left[\left(t_2 -\frac{r_2}{c} \right)-\left(t_1-\frac{r_1}{c} \right)\right]}\Theta\left[\left(t_2 - \frac{r_2}{c} \right) - \left(t_1 -\frac{r_1}{c} \right) \right]
\label{psi1D}
\end{multline}
where $\eta$ is some constant given in the Appendix~\ref{Ap:2}.
Under the assumption of equivalent relative phases between the short and the long paths for both photons,  $\phi_T$, see Eq.~\eqref{eq:assume_phaseT}, we then derive the second-order coherence function $G^{(2)}(\tau)$. In the case $\tau < -T $ we find
\begin{equation}
    G^{(2)}(\tau) = 0.
\end{equation}
If $ -T \leq \tau < 0 $, we have
\begin{equation}
     G^{(2)}(\tau) = \frac{\eta^2}{4 \Gamma_a} e^{-\Gamma_b \left(\tau+T\right)}
\end{equation}
while for $0 \leq \tau < T$ 
\begin{multline}
     G^{(2)}(\tau) = \frac{\eta^2}{2 \Gamma_a} e^{-\Gamma_b \tau} \left[1+ \frac12 e^{-\Gamma_b T}\right. \\
     \left. + e^{-\Gamma_a \frac{T}{2}}\cos\left(2\phi_T \right)  + \left(e^{-\Gamma_b \frac{T}{2}} + e^{-\left(\Gamma_a +\Gamma_b\right) \frac{T}{2}}\right)\cos\left(\phi_T\right) \right]
\end{multline}
and for $ \tau\geq T$
\begin{multline}
     G^{(2)}(\tau) = \frac{\eta^2}{2 \Gamma_a} e^{-\Gamma_b \tau} \left[1+ \frac12\left( e^{-\Gamma_b T} +  e^{\Gamma_b T}\right) + e^{-\Gamma_a \frac{T}{2}} \right. \\ \left. + e^{-\Gamma_a \frac{T}{2}}\cos\left(2\phi_T \right) + \left( 1 +  e^{-\Gamma_a \frac{T}{2}}\right)\left( e^{-\Gamma_b\frac{T}{2}} + e^{\Gamma_b\frac{T}{2}} \right)\cos\left(\phi_T \right) \right].
\end{multline}
The numerical and analytical results are compared in Fig.~\ref{fig:Benchmark} where we performed our numerical calculations up to a time $\Gamma_a t_\text{max} = 10$ to ensure that the 3LS has decayed to the ground state and two photons have been emitted. We see that both solutions agree perfectly which confirms the validity of our numerical approach. Furthermore, as illustrated in the inset, we observe the characteristic oscillation of the height of the central peak at $\tau = 0$ with the relative phase between the short and the long paths to the detectors, $\phi_T$. In the limit $\Gamma_a T \ll \Gamma_b T$, three separate peaks appear and the height of the central peak as a function of the phase $\phi_T$, $G^{(2)}_0(\phi_T)$, behaves as
\begin{equation}
    G^{(2)}_0(\phi_T) = \frac{\eta^2}{4\Gamma_a}\left[1 + e^{-\Gamma_a \frac{T}{2}} \cos(2\phi_T) \right]
\end{equation}
so that it takes on its maximal value at $\phi_T=0$ and its minimal value at $\phi_T=\pi/2$ while the height of the side peaks at $\tau = -T$ and $\tau=T$ does not change. This is due to the fact that we can clearly attribute the side peak at $\tau = -T$ to the first photon taking the long, the second photon taking the short path and the side peak at $\tau = T$ to the opposite case of the first photon taking the short, the second photon taking the long path. The central peak at $\tau=0$ originates from the two photons either both taking the short or both taking the long paths. The ignorance of the actual paths the photons take to the detectors, thus, gives rise to interference.
The visibility of this interference effect is defined as
\begin{equation}
    V \equiv \frac{G^{(2)}_0(0) - G^{(2)}_0(\frac{\pi}{2})}{G^{(2)}_0(0) + G^{(2)}_0(\frac{\pi}{2})}
\end{equation}
and depends crucially on $\Gamma_a T$ and $\Gamma_b T$ as shown in Fig.~\ref{fig:vis_noFB}. 

\section{Controlling the visibility of the $G^{(2)}$ interference}
\label{sec:results}

Implementing a coherent time-delayed feedback mechanism in our system allows us to modify the decay of the 3LS and, thus, opens up the possibility to control the time of emission of the photons. In this Section, we study the effect of feedback on the dynamics as well as on the second-order coherence function, $G^{(2)}(\tau)$, focussing on the case where only the first decay channel, that is, the $\ket{a} \leftrightarrow \ket{b}$ transition of the 3LS with transition frequency $\omega_{ab}$, is subjected to feedback.

\subsection{Dynamics}

A photon emitted into the feedback channel is reflected on the mirror and fed back to the 3LS after a delay time $\tau_\text{FB}$ as illustrated in Fig.~\ref{fig:FransonSetup}. Due to the interference of the feedback signal with the emission from the 3LS at that time, the usual Wigner-Weißkopf decay is modified. 
The effect of the feedback signal depends on the phase a photon acquires during one feedback round trip, $\phi_\text{FB}\equiv \omega_{ab}\tau_\text{FB}$.
The decay of the upper state of the 3LS is fastest if a feedback phase $\phi_\text{FB} = (2n+1)\pi$, $n \in \mathbb{N}$, is implemented while the decay is slowed down maximally for a feedback phase $\phi_\text{FB} = 2n\pi$. In a two-level system, such a phase can completely stop the excitation decay after a transient time so that a finite excitation probability of the emitter in the long-time limit is possible while in our system with feedback on the $\ket{a} \leftrightarrow \ket{b}$ transition, the 3LS always decays to the ground state emitting two photons. 

In Sec.~\ref{sec:Model}, we saw that the visibility of the $G^{(2)}$ interference for the conventional Franson interferometer depends strongly on the decay rates of the upper and middle state of the 3LS and is highest for a slowly decaying upper state and a fast decaying middle state.
It is, thus, interesting to implement a feedback mechanism which slows down the excitation decay from the upper level in relation to the decay of excitation from the middle state to study the potential of feedback to enhance the visibility of the $G^{(2)}$ interference.

\begin{figure}
    \centering
    \includegraphics[]{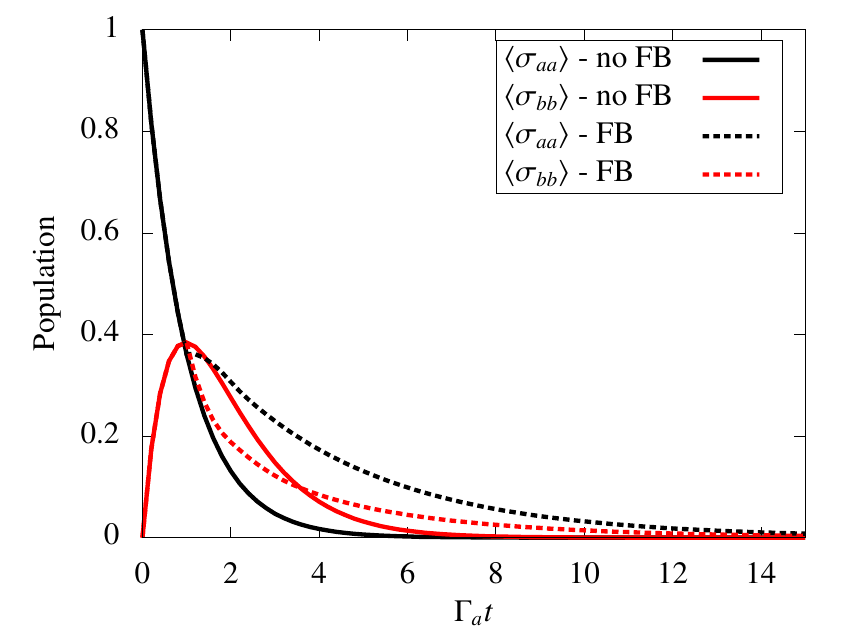}
\caption{(Color online) Dynamics of the population of the upper level ($\braket{\sigma_{aa}}$) and the middle level ($\braket{\sigma_{bb}}$) of a 3LS with $\Gamma_a = \Gamma_b$ without feedback (no FB) and with feedback at $\Gamma_a \tau_\text{FB}=1$ (FB), $\Delta t=0.05/\Gamma_a$.}
    \label{fig:dyn_fb_first_ch}
\end{figure}

In Fig.~\ref{fig:dyn_fb_first_ch}, we exemplarily consider a 3LS with $\Gamma_a  = \Gamma_b$ and compare the dynamics for a system without feedback and a system in which feedback with a phase $\phi_\text{FB}= 2\pi n$ is implemented for the first transition of the 3LS at $\Gamma_a \tau_\text{FB} = 1$. With feedback, the decay of excitation from the upper to the middle state of the 3LS is slowed down in comparison to the case without feedback. The modified decay to the middle state also affects the decay from the middle to the ground state. Nevertheless, the implemented feedback mechanism slows down the decay of the upper state in relation to the decay of the middle state considerably. 
For the MPS calculations in this case, as well as throughout the rest of this work, we did not limit the bond dimension when performing singular value decompositions. This corresponds to keeping all singular values which is numerically feasible since we consider only two excitations and limited delay times.

In the following, if the system is subjected to feedback, we concentrate on this case of $\phi_\text{FB} = 2\pi n$ where the decay of the upper state can be slowed down maximally.

\subsection{Second-order coherence function}

\begin{figure}
    \centering
    \includegraphics[]{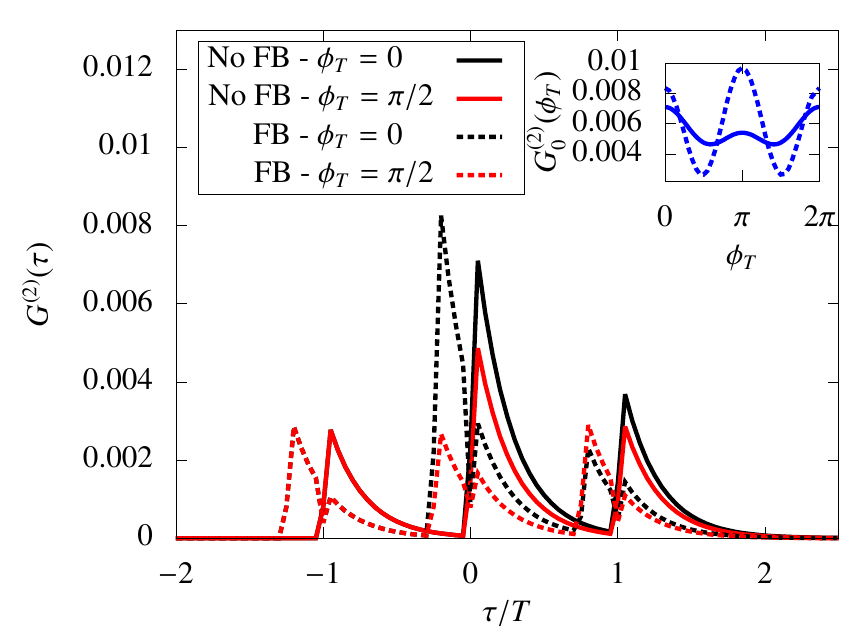}
\caption{(Color online) Second-order coherence function for a 3LS with $\Gamma_a T = \Gamma_b T = 4$ without feedback (no FB) and with feedback at $\Gamma_a\tau_\text{FB} = 1$ (FB), $\Delta t=0.2/\Gamma_a$, $\Gamma_a t_\text{max} = 20$. (Inset) Oscillation of the height of the central peak with the delay phase $\phi_T$ without feedback (solid line) and with feedback (dashed line).}
    \label{fig:g2_fb_first_ch_1}
\end{figure}

After discussing in which way feedback allows to control the dynamics of the 3LS, we study its impact on the second-order coherence function.
For the system the dynamics of which we presented in Fig.~\ref{fig:dyn_fb_first_ch}, we compare $G^{(2)}(\tau)$ without and with feedback in Fig.~\ref{fig:g2_fb_first_ch_1}. To ensure that the 3LS has completely decayed for all feedback phases, the time evolution has been performed up to $\Gamma_a t_\text{max} = 20$.
Since in this system $\Gamma_a = \Gamma_b$, we are not in the parameter regime in which a high visibility of the interference in the second-order coherence function can be expected. Indeed, without feedback, we find a visibility of the $G^{(2)}$ interference of $V=0.19$. With feedback, the peaks are split and shifted on the x-axis by $-\tau_\text{FB}$. For a delay phase $\phi_T = 0$ between the short and the long paths to the detectors, the height of the central peak at $\tau = -\tau_\text{FB}$ is increased in comparison to the case without feedback and reduced for a delay phase $\phi_T = \pi/2$ which results in a visibility of $V= 0.51$. This corresponds to an increase of 168\% in comparison to the case without feedback and even goes beyond the classical limit where a maximum visibility of $V=0.5$ is possible \cite{Ou1990,Franson1991a,Kwiat1993}.
Note that the violation of the classical limit does not imply a violation of a Bell-type inequality which would require a higher visibility but demonstrates and quantifies the non-classical nature of the light \cite{Aerts1999,Jogenfors2014}.
In the inset of Fig.~\ref{fig:g2_fb_first_ch_1}, the oscillation of the height of the central peak with the delay phase $\phi_T$ is shown and we see that the oscillation is significantly more pronounced in the case with feedback (dashed line) than without feedback (solid line).

\begin{figure}
    \centering
    \includegraphics[]{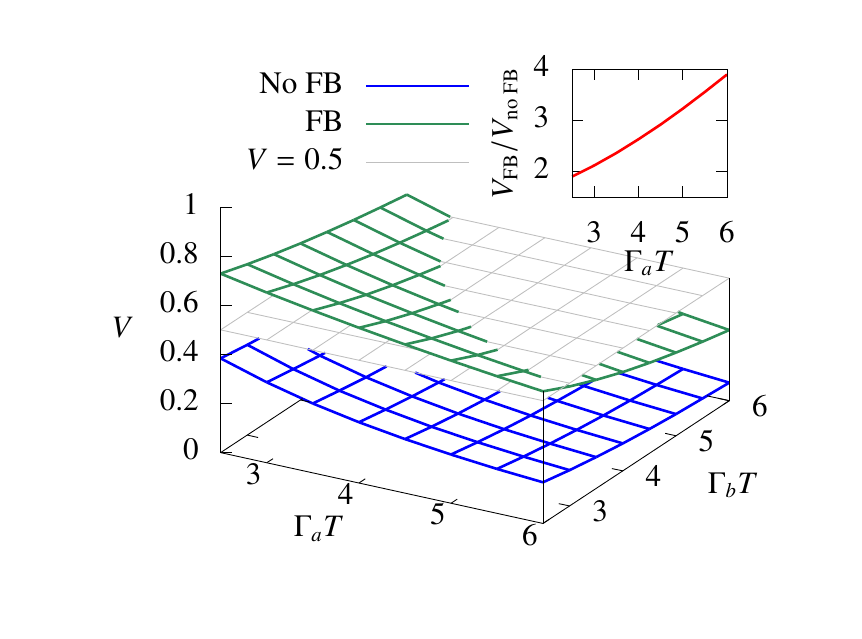}
\caption{(Color online) Comparison of the visibility of the $G^{(2)}$ interference as a function of $\Gamma_a T$ and $\Gamma_b T$ without feedback (no FB) and with feedback at $\tau_\text{FB} = T/4$ (FB), $\Delta t = 0.05 T$. (Inset) Enhancement of the visibility of the $G^{(2)}$ interference with feedback ($V_\text{FB}$) in relation to the visibility without feedback ($V_\text{no FB}$) as a function of $\Gamma_a T$ for $\Gamma_b = \Gamma_a$.}
    \label{fig:vis}
\end{figure}

The potential of feedback to increase the visibility of the $G^{(2)}$ interference depends on the system parameters. In Fig.~\ref{fig:vis}, the visibility as a function of $\Gamma_a T$ and $\Gamma_b T$ for a system subjected to feedback at $\tau_\text{FB} = T/4$ is shown and compared to the visibility without feedback. To ensure that two photons have been emitted from the 3LS, we have adapted the simulation time $t_\text{max}$ according to the decay rates $\Gamma_a$ and $\Gamma_b$ and checked for the complete decay of the excitation in the 3LS.
Here, we focus on the parameter regime of relatively large $\Gamma_a T$ and $\Gamma_b T$ where without feedback a visibility well below the classical limit of $V=0.5$ is found. With feedback, the visibility is raised above this threshold in much of the considered parameter area. While the visibility is enhanced remarkably for all considered parameter values, the impact of the feedback increases with $\Gamma_a T$ as illustrated in the inset of Fig.~\ref{fig:vis}. Furthermore, the enhancement is largely independent of $\Gamma_b T$. This can be attributed to the fact that it is the $\ket{a}\leftrightarrow \ket{b}$ transition with decay rate $\Gamma_a$ that is subjected to feedback in our setup. In addition to that, the influence of $\Gamma_a T$ on the visibility is inherently larger than the influence of $\Gamma_b T$ as illustrated in Fig.~\ref{fig:vis_noFB}.

\section{Conclusion and outlook}
\label{sec:Conclusion}

We examined the possibility to control the visibility of the two-photon interference in the Franson interferometer via a coherent time-delayed feedback mechanism.
With a 3LS in ladder configuration as the source of the two photons, in the conventional Franson setup, a high visibility can be found for an upper state decaying slowly in comparison to the decay of the middle state. If the 3LS is implemented as a biexciton-exciton cascade in a semiconductor quantum dot, this is difficult to obtain since the biexciton state usually has a significantly lower lifetime than the exciton state.
To obtain a high visibility in other, more realistic parameter regimes as well, we suggested slowing down the decay of the upper state by implementing a feedback channel. This complements approaches such as Purcell enhancement or suppression in micro- and nanocavities where the individual control of the transitions is difficult to realize.
Modeling the non-Markovian dynamics within the MPS framework, we studied the impact of the feedback on the second-order coherence function and found that feedback has the potential to enhance the visibility significantly. The faster the decay of the upper state, the stronger the effect of the feedback.

Since the visibility of the two-photon interference in the Franson interferometer indicates the energy-time entanglement of the photons, we conclude that coherent feedback opens up the possibility to control and, in particular, enhance the energy-time entanglement of the photons.

\section*{Acknowledgements}
The authors gratefully acknowledge the support of the Deutsche Forschungsgemeinschaft (DFG) through the project B1 of the SFB 910 (Project No. 163436311) and the project RE2974/18-1.

\begin{appendices}
\section{Stroboscopic time evolution operator}
\label{Ap:0}

We derive the stroboscopic time evolution operator from time step $k$ to $k+1$, $k \in \mathbb{N}$, $t_{k+1}-t_k = \Delta t$,
\begin{equation}
    U_k = \exp\left[-\frac{i}{\hbar}\int_{t_k}^{t_k+1}\mathrm{d}t' \mathcal{H}''(t') \right]
    \label{eq:Ap_Uk}
\end{equation}
where no explicit time ordering is necessary if time is discretized into sufficiently small steps $\Delta t$.
Here, $\mathcal{H}''(t)$ is obtained from the Hamiltonian of the system in the rotating frame, $\mathcal{H}'(t)$, given in Eq.~\eqref{eq:Ham_rot} of the main text with feedback implemented at delay time $\tau_\text{FB}$ for the first channel, that is, with $g_a(\omega) = g_a\sin\left(\omega \tau_\text{FB}/2\right)$, $g_b(\omega) = g_b$, after a time-independent phase shift via the unitary transformation
\begin{equation}
    \mathcal{H}''(t) = U^\dagger \mathcal{H}'(t)U \text{,}\quad U = \exp\left[i \int\mathrm{d}\omega r^{\dagger (1)}_\omega r^{(1)}_\omega \frac{\omega \tau_\text{FB}}{2}  \right].
\end{equation}
%
%
%
Furthermore, we perform a transformation from the frequency to the time domain introducing the time-dependent noise operators
\begin{equation}
    r^{\dagger (i)}_t = \frac{1}{\sqrt{2\pi}}\int \md \omega r_\omega^{\dagger (i)} e^{i\left(\omega-\omega_i\right)t},
\end{equation}
$i \in \{1,2\}$, with $\omega_1 = \omega_{ab}$, $\omega_2 = \omega_b$ which obey 
\begin{equation}
    \left[r_t^{(i)}, r_{t'}^{\dagger (j)} \right] = \delta_{ij}\delta(t-t').
\end{equation}
While the operators $r_\omega^{\dagger (i)}$ describe the creation of a photon with frequency $\omega$ in the reservoir $i$, as their fourier transforms the operators $r_t^{\dagger(i)}$ model the creation of a photon at time $t$ in the respective reservoir. This way, the Hamiltonian $\mathcal{H}''(t)$ can be expressed as
\begin{multline}
    \mathcal{H}''(t) = i \hbar \sqrt{\Gamma_a} \left[\sigma_+^{(1)} \left[r^{(1)}_t - r^{(1)}_{t-\tau}e^{i \omega_{ab} \tau_\text{FB}} \right] -\text{H.c.}\right] \\
    + \hbar \sqrt{\Gamma_b} \left[\sigma_+^{(2)} r^{(2)}_t+ \text{H.c.}\right]
\end{multline}
with the decay rates $\Gamma_a \equiv g_a^2 \pi/2$ and $\Gamma_b \equiv 2\pi g_b^2$.
Introducing the noise increments
\begin{equation}
    \Delta R_k^{\dagger (i)} = \int_{t_k}^{t_{k+1}}\md t r^{\dagger (i)}_t
\end{equation}
which describe the creation of a photon in the reservoir $i$ in time step $t_k$, we discretize the time and
the stroboscopic time evolution operator from Eq.~\eqref{eq:Ap_Uk}, thus, takes the form
\begin{multline}
    U_k = \exp\left\{ \sqrt{\Gamma_a} \left[\sigma_+^{(1)} \left(\Delta R^{(1)}_k - \Delta R^{(1)}_{k-m}e^{i \omega_{ab} \tau_\text{FB}} \right) -\text{H.c.}\right] \right.  \\
    \left.-i \sqrt{\Gamma_b}  \left[\sigma_+^{(2)} \Delta R^{(2)}_k + \text{H.c.}\right] \right\}
\end{multline}
where $\tau_\text{FB} = m \Delta t$.

\section{Two-photon state emitted from the 3LS}
\label{Ap:1}

We use the Hamiltonian in the rotating frame, $\mathcal{H}'(t)$, given in Eq.~\eqref{eq:Ham_rot} of the main text where without feedback it holds that $g_a(\omega) = g_a$ and $g_b(\omega)= g_b$ so that
\begin{multline}
    \mathcal{H}'(t) = \hbar \int \mathrm{d}\omega g_a \left(\sigma_+^{(1)}r_\omega^{(1)}e^{i(\omega_{ab}-\omega)t} +\text{H.c.} \right) \\
    + \hbar \int \mathrm{d}\omega g_b \left(\sigma_+^{(2)}r_\omega^{(2)}e^{i(\omega_b-\omega)t} +\text{H.c.} \right)
\end{multline}
to derive the coefficients of the combined state of the 3LS and the reservoir in the two-excitation limit
\begin{multline}
    \ket{\psi(t)} = c_a(t)\ket{a,0} + \int\mathrm{d}\omega c_{b,\omega}(t)\ket{b,1_\omega} \\
    + \int\hspace{-0.5em} \int \mathrm{d}\omega \mathrm{d}\omega' c_{c,\omega,\omega'}(t)\ket{c,1_\omega,1_{\omega'}}.
\end{multline}
Here, the first term on the right-hand side of the equation describes the 3LS in the upper state and no photons in the reservoir, the second term represents the 3LS in the middle state and a photon with frequency $\omega$ in the reservoir while the third term refers to the 3LS in the ground state and two photons in the reservoir, one with frequency $\omega$, the other with frequency $\omega'$. With the Schrödinger equation, we find
\begin{align}
    \dot{c}_a(t) &= -i g_a \int\mathrm{d} \omega e^{i\left(\omega_{ab}-\omega\right)t}c_{b,\omega}, \\
    \dot{c}_{b,\omega}(t) &= -i g_a e^{-i\left(\omega_{ab}-\omega\right)t}c_a(t) \notag \\
    &\qquad\qquad - i \int \mathrm{d}\omega' g_b e^{i\left(\omega_b-\omega'\right)t} c_{c,\omega,\omega'}(t), \\
    \dot{c}_{c,\omega,\omega'}(t) &= -i g_b e^{-i\left(\omega_b - \omega'\right)t} c_{b,\omega}(t).
\end{align}
Using the Weisskopf-Wigner theory of spontaneous emission \cite{Scully1997}, we can approximate the decay of an atomic state in a continuum of photon modes via
\begin{align}
-ig_a\int\mathrm{d}\omega e^{i\left(\omega_{ab}-\omega\right)t}c_{b,\omega}(t) &= -\frac{\Gamma_a}{2}c_a(t), \\
-ig_b\int\mathrm{d}\omega' e^{i\left(\omega_{b}-\omega'\right)t}c_{c,\omega,\omega'}(t) &= -\frac{\Gamma_b}{2}c_{b,\omega}(t).
\end{align}
where $\Gamma_a$ and $\Gamma_b$ are the decay rate of the upper state $\ket{a}$ and the middle state $\ket{b}$, respectively.
As a consequence, starting in the upper state, that is, with the initial conditions $c_a(0)=1, c_{b,\omega}(0) = c_{c,\omega,\omega'}(0)=0$ for all $\omega$, $\omega'$, we obtain
\begin{align}
    c_a(t) &= e^{-\frac{\Gamma_a}{2}t}, \\
    c_{b,\omega}(t) &= \frac{-ig_a}{i\left(\omega-\omega_{ab}\right) -\left(\frac{\Gamma_a}{2} - \frac{\Gamma_b}{2}\right)}  \left(e^{-i\left(\omega_{ab}-\omega\right)t-\frac{\Gamma_a}{2}t} - e^{-\frac{\Gamma_b}{2}t}\right), \\
    c_{c,\omega,\omega'}(t) &= \frac{-ig_ag_b}{i\left(\omega-\omega_{ab}\right)-\left(\frac{\Gamma_a}{2} -\frac{\Gamma_b}{2}\right)}\notag \\
    & \times\int_0^t\mathrm{d}t'\left( e^{-i\left(\omega_{a}-\omega-\omega'\right)t'-\frac{\Gamma_a}{2}t'} - e^{-i\left(\omega_b-\omega'\right)t'-\frac{\Gamma_b}{2}t'} \right).
\end{align}
Since we want to study two-photon interference effects, we are interested in the long-time limit where the 3LS has completely decayed to the ground state and two photons have been emitted. In this limit, the coefficients $c_a(t)$ and $c_{b,\omega}(t)$ have decayed to zero and omitting the 3LS contribution we obtain the two-photon state
\begin{multline}
 \ket{\Psi} = \hspace{-0.5em} \bigintsss\hspace{-1em}\bigintsss \hspace{-0.5em} \mathrm{d}\omega \mathrm{d}\omega' \\ \times \frac{ - g_a g_b}{\left[i \left(\omega +\omega'- \omega_{a} \right) - \frac{\Gamma_a}{2} \right]\left[i \left( \omega'-\omega_{b}\right) -\frac{\Gamma_b}{2} \right]} \ket{1_{\omega}, 1_{\omega'}}.
\end{multline}

\section{Two-photon probability amplitude}
\label{Ap:2}

To derive the two-photon probability amplitude
\begin{equation}
\Psi_{r_1,r_2}(t_1,t_2) = \bra{0}E^{(+)}_{2,r_2}(t_2)E^{(+)}_{1,r_1}(t_1)\ket{\Psi}
\end{equation}
of the first photon taking the path $r_1$ to the first detector and the second photon reaching the second detector via the path $r_2$ with $r_1$, $r_2 \in \{S,L\}$  we use the two-photon state given in Eq.~\eqref{eq:two-photState} of the main text and the positive frequency part of the electric field operator \cite{Scully1997}
\begin{equation}
E^{(+)}_{i,r}\left(t\right) = \sum_{\omega} \mathcal{E}(\omega) r^{(i)}_\omega e^{-i \omega\left(t- \frac{r}{c}\right)}, \quad \mathcal{E}(\omega) = \sqrt{\frac{\hbar \omega}{2 \epsilon_0 V}} \label{E+_1D}
\end{equation}
with vacuum permittivity $\epsilon_0$ and quantization volume $V$.
Switching from a sum to an integral, the operator can be expressed as
\begin{equation}
   E^{(+)}_{i,r}\left(t\right) = \int \mathrm{d}\omega \sigma(\omega) \mathcal{E}(\omega)r^{(i)}_\omega e^{-i\omega\left(t-\frac{r}{c} \right)} 
\end{equation}
where $\sigma(\omega)$ is the density of states.
Approximating $\sigma(\omega)$ and $\mathcal{E}(\omega)$ with their values at the atomic resonances, we obtain
\begin{multline}
\Psi_{r_1,r_2}(t_1,t_2) = -g_{a}g_{b}\sigma\left(\omega_{ab}\right)\sigma\left(\omega_{b}\right)\mathcal{E}\left(\omega_{ab}\right)\mathcal{E}\left(\omega_{b}\right) \\
\times \bigintsss\hspace{-1em}\bigintsss \hspace{-0.5em} \mathrm{d}\omega \mathrm{d}\omega' \frac{ e^{-i\omega\left(t_1-\frac{r_1}{c}\right)}e^{-i\omega'\left(t_2-\frac{r_2}{c}\right)}}{\left[i \left(\omega +\omega'- \omega_{a} \right) - \frac{\Gamma_a}{2} \right]\left[i \left( \omega'-\omega_{b}\right) -\frac{\Gamma_b}{2} \right]}.
\end{multline}
The evaluation of the integrals using the residue theorem finally yields
\begin{multline}
\Psi_{r_1,r_2}(t_1,t_2) =  \eta e^{-\left( i\omega_{a}+\frac{\Gamma_a}{2}\right)\left(t_1 - \frac{r_1}{c} \right)} \Theta\left(t_1 - \frac{r_1}{c} \right) \\
\times e^{-\left(i\omega_{b} + \frac{\Gamma_b}{2}\right)\left[\left(t_2 -\frac{r_2}{c} \right)-\left(t_1-\frac{r_1}{c} \right)\right]}\Theta\left[\left(t_2 - \frac{r_2}{c} \right) - \left(t_1 -\frac{r_1}{c} \right) \right]
\end{multline}
with $\eta \equiv 4 \pi^2 g_{a}g_{b}\sigma\left(\omega_{ab}\right)\sigma\left(\omega_{b}\right)\mathcal{E}\left(\omega_{ab}\right)\mathcal{E}\left(\omega_{b}\right)$.

\end{appendices}

\bibliographystyle{apsrev4-2}
\bibliography{BibCollection}
\clearpage

\end{document}